# BREVE HISTORIA DE LA ASTROBIOLOGÍA EN ARGENTINA

## Ximena C. Abrevaya[1,2,*]

[1]*Instituto de Astronomía y Física del Espacio (IAFE), CONICET – UBA. Ciudad Autónoma de Buenos Aires, Argentina.*

[2]*Facultad de Ciencias Exactas y Naturales, Universidad de Buenos Aires. Ciudad Autónoma de Buenos Aires, Argentina.*

*Ximena C. Abrevaya. Pabellón IAFE, Ciudad Universitaria, CC 67, Suc. 28, 1428. Ciudad Autónoma de Buenos Aires, Argentina. E-mail: abrevaya@iafe.uba.ar

## Introducción

La astrobiología o exobiología es un área de la ciencia relativamente nueva que indaga sobre las posibilidades de encontrar vida en otros lugares del universo. Esto no sólo incluye la exploración de planetas cercanos o distantes a la Tierra, sino también la exploración de nuestro propio planeta en el estudio del origen y evolución de la vida. Debido a la complejidad de las preguntas que intenta responder, la astrobiología se nutre de diversas áreas de la ciencia, como la física, la química, la astronomía, la geología, entre otras, por lo que es considerada un área interdisciplinaria o transdisciplinaria. Algunas de estas preguntas también han surgido desde estas áreas y han convergido en lo que hoy día es la astrobiología. Incluso muchas de estas preguntas también han sido formuladas en la antigüedad, aunque sin el marco científico y técnico con el que contamos hoy en día. Es el caso de civilizaciones como las de la antigua Grecia, entre otras, que han planteado cuestiones directamente relacionadas a las posibilidades de encontrar vida en otros mundos. Por lo tanto, muchos de los cuestionamientos hoy son parte de la astrobiología anteceden al desarrollo de la astrobiología en sí misma.

Este ha sido el camino que ha llevado a desarrollar a la astrobiología como una ciencia joven a nivel mundial. Muchos países de Latinoamérica, como Argentina, no han sido la excepción a ello.

## Primeros comienzos

En particular, si buscamos en Argentina los orígenes del interés por explorar las preguntas que la astrobiología hoy plantea, nos encontramos con que los primeros antecedentes relacionados a cuestiones "astrobiológicas" se han dado en los albores de la República Argentina, en una época que el historiador Miguel de Asúa ha llamado la "primavera científica de la década de 1820" (de Asúa M., 2011).

En esa primavera científica, aparece como protagonista el trabajo de un político, militar y médico, Manuel Moreno (Buenos Aires, 1781-1852), hermano del prócer argentino Mariano Moreno (Fig 1). Entre sus cargos se destacaron, el de oficial en la Real Hacienda del Virreinato, el de subteniente habiendo luchado en la defensa de Buenos

Aires contra los ingleses durante las invasiones, y el de oficial mayor de la secretaría de Estado en 1813. Más tarde, Moreno emigró a Estados Unidos donde estudió medicina en la Universidad de Maryland, Baltimore. Luego de recibir el título de Bachelor y doctor en medicina regresó a Buenos Aires en 1821.

Precisamente en esos años, entre 1821 y 1823, también se crearon varias instituciones científicas importantes, como la Universidad de Buenos Aires, la Academia Nacional de Medicina, la Sociedad Literaria y la Sociedad de Ciencias Físico-Matemáticas. La Universidad de Buenos Aires no sólo era una universidad en los términos actuales sino que también funcionaba como "ministerio" y estaba a cargo de toda la enseñanza, incluyendo la de nivel medio. Estaba dividida en seis departamentos: de primeras letras, de estudios preparatorios (enseñanza media), de medicina, de ciencias exactas, de ciencias sagradas y de jurisprudencia.

Precisamente, la cátedra de química experimental del Departamento de Estudios Preparatorios estuvo a cargo de Manuel Moreno, quien también presidió la Academia Nacional de Medicina (de Asúa M., 2011). Las evidencias que vinculan a Moreno con temas relacionados a lo que hoy conocemos como astrobiología tienen que ver con un periódico que existía en aquel entonces llamado "La Gazeta de Buenos Ayres" (Fig. 2), creado por su hermano Mariano Moreno, que publicaba los documentos y resoluciones oficiales y decretos gubernamentales pero que además ofrecía noticias de actualidad tanto del extranjero como locales. Según relata Guillermo Lemarchand en su libro "Astrobiología: del Big Bang a las civilizaciones", Moreno discutió en La Gazeta acerca del origen de la vida en la Tierra y en la posible existencia de vida en otros planetas (Lemarchand G., 2010)

En el mismo libro, Lemarchand destaca también a Vicente Fidel López (Buenos Aires, 1815-1903) historiador, abogado y político, hijo del compositor del himno nacional argentino. López, quien también fue rector de la UBA, habría discutido la obra del científico y explorador alemán, Alexander von Humboldt, llamada "Kosmos" (Fig 3), un tratado de cinco volúmenes sobre ciencia y naturaleza. En el primer volumen von Humboldt describe la naturaleza física del espacio exterior y de la Tierra, y López, basándose en este libro habría especulado en una serie de cartas acerca del vulcanismo y de la capacidad de este proceso de generar vida en otros mundos (Lemarchand G., 2010).

Treinta años después de que Vicente Fidel López muriera, nacía Carlos Manuel Varsavsky (1933-1983), físico y astrofísico argentino (Fig. 4). Varsavsky emigró a Estados Unidos luego de haber cursado sus estudios de nivel medio en donde posteriormente obtuvo una licenciatura y un máster en Ingeniería Física por la Universidad de Colorado. Luego se doctoró en astronomía en la Universidad de Harvard en 1959. Su trabajo de tesis se relacionó a transiciones atómicas de interés astrofísico y durante décadas fue una obra de referencia.

Un año antes de que Varsavksy obtuviera su doctorado, en 1958, se creaba en la Argentina la Comisión de Astrofísica y Radioastronomía (CAR), luego de que se instalara en la Facultad de Agronomía de la Universidad de Buenos Aires (UBA) un interferómetro solar en 86 MHz. La CAR estaba integrada por los Dres. Félix Cernuschi y Enrique Gaviola (presidente) y por el Ing. Humberto Ciancaglini y había surgido motivada por el viaje a la Argentina del Dr. Merle Tuve, de la Carnegie Institution of Washington(CIW), EEUU. La CIW tenía entre sus intenciones, la creación del observatorio que se llamaría "Carnegie Southern Station for Radio Astronomy", debido a la necesidad de obtener señales desde latitudes australes. En la construcción del inteferómetro solar habían intervenido varios estudiantes, entre ellos Fernando Raúl Colomb (1939-2008) y Valentín Boriakoff (1938-1999), que luego tendrían un protagonismo fundamental en el tema de la búsqueda de civilizaciones inteligentes. Si bien el interferómetro no condujo a resultados de interés científico, cumplió con el objetivo inicial del Dr. Tuve, que fue el de despertar el interés en la radioastronomía en Argentina (Reseña histórica del IAR, s.f.; Bajaja, E., s.f).

Poco tiempo después de que la CAR fuese creada, en 1960, Carlos Varsavsky regresa a Buenos Aires donde se incorporó al grupo de Astrofísica de la UBA. También obtiene un cargo como profesor titular de física en la Facultad de Ciencias Exactas y Naturales de dicha universidad, en el que se desempeñó hasta 1966.

Debido al auge de la radioastronomía en el país alrededor de esa época, exactamente en 1962, se crea el Instituto Argentino de Radioastronomía (IAR) como iniciativa entre varias instituciones: el Consejo Nacional de Investigaciones Científicas y Técnicas (CONICET), la Comisión de Investigaciones Científicas de la Provincia de Buenos Aires (CIC), la Universidad Nacional de La Plata (UNLP) y la UBA.

Así, en 1963 se inicia la construcción de la antena parabólica de 30 metros de diámetro que sería parte del IAR, en conjunto con obras civiles en la locación elegida, muy cerca de la ciudad de La Plata, provincia de Buenos Aires. La CIW de EEUU colaboró también con partes de la primera antena. Este sería en aquel entonces el radiotelescopio más grande de Sudamérica. Entre las funciones del IAR estaban las de "promover y coordinar la investigación y desarrollo técnico de la radioastronomía y colaborar en la enseñanza". Así científicos e ingenieros viajaron al exterior para perfeccionar sus conocimientos y adquirir experiencia en técnicas de observación de la línea de 21cm (correspondiente al hidrógeno neutro, el elemento más abundante en el universo), que fue detectada en forma experimental por primera vez en 1965 (Reseña histórica del IAR, s.f.).

El 26 de marzo de 1966 se inaugura oficialmente el IAR con la finalización de la construcción de la primera antena (Fig. 5). Varsavsky se convierte en su primer director, desempeñando un papel fundacional.

Dos años después, en 1968, Varsavsky publica el libro "Vida en el Universo" (Fig. 6). Este fue el primer libro de un científico argentino, y posiblemente el primer libro de un científico latinoamericano, que trata el tema de la búsqueda de vida inteligente en el cosmos. El texto tiene un estilo divulgativo de excelencia y hoy día es uno de los referentes en el terreno de la comunicación pública de la ciencia.

La temática del libro de Varsavsky junto con el hecho de que el que estuviese involucrado en el área de la radioastronomía no es casual. Eran épocas en las que la llamada "Ecuación de Drake" estaba en pleno auge, dado que había sido formulada sólo unos años antes por el mismo Frank Drake, en 1961. No sólo esto, sino que en verdad ya a partir de 1959 los radiotelescopios empiezan a cobrar un protagonismo adicional que surge a partir de un artículo en la revista Nature. El artículo, publicado por dos físicos, Giuseppe Cocconi y Phillip Morrison de la Universidad de Cornell, EEUU, titulado "Searching for interstellar communications", propone que la línea de 21 cm detectada por los radiotelescopios podría ser utilizada en la comunicación con civilizaciones extraterrestres. Posteriormente, luego de que Cocconi y Morrison publicaran su artículo, Frank Drake hace la primera búsqueda de estas señales de "vida inteligente" en el contexto del Proyecto "Ozma" en los años 60. El proyecto se desarrolló en el National Radio Astronomy Observatory (NRAO) en Green Bank, West Virginia, EEUU.

**El proyecto SETI en Argentina**

En Argentina, el IAR comienza a albergar sus primeros estudiantes alrededor de los años 60, entre ellos Fernando Colomb, que había participado de la construcción del interferómetro solar de parte de la CAR y se desempeñó como ayudante técnico en ambas ocasiones, en el caso del IAR entre 1963 y 1965, año en el que obtuvo el título de licenciado en Ciencias Físicas de la UBA.
Ya en 1966 Colomb viaja al mismo observatorio radioastronómico donde Frank Drake hace la primera búsqueda de señales inteligentes, el NRAO, y allí trabaja como asistente de investigación.

Otro de los estudiantes que se encontraba de manera contemporánea a Colomb en el IAR fue Valentín Boriakoff, ingeniero electromecánico con orientación en electrónica. Boriakoff estuvo dedicado a tareas de ingeniería electrónica en el observatorio, antes de emigrar a EEUU y al NRAO tal como lo hiciera Colomb. En 1973 obtuvo su doctorado en la Universidad de Cornell, habiendo trabajado con Frank Drake en observaciones de radiofrecuencia de púlsares.

En su haber Boriakoff tiene otro importante logro relacionado a la misión Voyager. Tal como relata Guillermo Lemarchand: "en 1977, Boriakoff resuelve la manera de registrar imágenes en formato $33^{1/2}$ r.p.m. para el disco interestelar de las naves Voyager I y II, colaborando de esta manera en el diseño de un mensaje interestelar destinado a una hipotética civilización extraterrestre". De hecho, Carl Sagan menciona a Boriakoff en su libro "The sounds of Earth" (traducido en su versión en español como "Murmullos de la Tierra"). Como nota de color adicional una foto de Boriakoff aparece como una de las imágenes incluídas en el disco de oro de las Voyager (Lemarchand G., 2010) (Fig. 7).

Ya comenzados los años 80', la Unión Astronómica Internacional (IAU) crea la comisión 51 dedicada a la Bioastronomía, aún en vigencia. Un científico uruguayo y nacionalizado argentino, participó de un simposio organizado en Boston por dicha comisión, Félix Mirabel. Graduado como licenciado y doctor astronomía por la Universidad Nacional de La Plata, Mirabel además obtuvo un título como doctor en filosofía por la UBA. Mirabel, uno de nuestros más destacados astrónomos, fue quien propuso en dicha reunión que se establezca un proyecto sobre búsqueda de señales de origen extraterrestre en Sudamérica (Mirabel I. F., 1984).

Posteriormente Mirabel trabajó en la búsqueda de señales de comunicación interplanetaria en el hoyo anti máser de la molécula de formaldehido con el radiotelescopio de 140 pies de NRAO. También participó de la organización de reuniones sobre la temática a su regreso a la Argentina a partir del año 1983 (Comunicación personal con el Dr. Mirabel).

De hecho, el Dr. Mirabel, además fue uno de los ponentes invitados en las "Primeras Jornadas Interdisciplinarias Sobre Vida Inteligente en el Universo", organizadas por un grupo de estudiantes de la Facultad de Ciencias Exactas y Naturales de la UBA en 1985 (Lemarchand G. A., 1986). Ese mismo grupo de estudiantes conformaba lo que se llamó la "Comisión de Astrofísica" que tenía como objetivo estimular los estudios de física aplicada al espacio debido a la que UBA no ofrece la carrera de Astronomía. Dicha comisión también publicó durante varios años una revista, llamada "Astrofísica", donde se trataban varios temas, incluyendo aquellos relacionados a la astrobiología (Fig 8).

En el grupo de estudiantes, el impulsor y organizador principal de la Comisión y de esas jornadas fue Guillermo Lemarchand, un joven estudiante de física en aquel entonces. Las jornadas, que duraron tres días, fueron un éxito. Involucró un total de unos 500 participantes que incluyeron estudiantes y público general y contó con invitados como Félix Mirabel, Robert Bruce Cow de la NASA y Fernando Raúl Colomb que para entonces se había convertido en director del IAR (Lemarchand, G. A. 1986; Reseña histórica del IAR, s.f.). Pero no sólo eso, las jornadas además facilitaron el acercamiento y apertura de la temática al país. Así fue como posteriormente se llegó a la firma de un convenio de cooperación entre el IAR y The Planetary Society para el establecimiento del programa SETI (Search for Extraterrestrial Intelligence) en Argentina. Este proyecto sería el primero y único del cono sur.

El 7 de octubre de 1986 la antenas del IAR hicieron las primeras observaciones del cielo en búsqueda de señales de civilizaciones que pudiesen provenir de estrellas cercanas (link página SETI). Las investigaciones estuvieron a cargo del Dr. Colomb, cuyo grupo de investigación estaba integrado por Guillermo Lemarchand y María Cristina Martín.

El trabajo que se estaba realizando en el IAR estimuló la organización de un seminario sobre aspectos "radioastronómicos, sociales y jurídicos de la detección de señales electromagnéticas de origen extraterrestre", que estuvo a cargo del jurista internacional Aldo Armando Cocca. El jurista argentino, quien estudió abogacía en la UBA, fue el primer argentino en realizar una tesis en Derecho Espacial (1953). Debido a esto se lo conoce como el "fundador del Derecho Espacial Argentino". Al año siguiente, crea el concepto de "Patrimonio Común de la Humanidad" aplicado a los cuerpos celestes, que luego se incluyeron en el "Acuerdo sobre la Luna", en la ONU (1979). También fue el ideólogo del Planetario de la Ciudad de Buenos Aires.

En 1988, Lemarchand viaja a la Universidad de Harvard donde funcionaba el programa SETI. El objetivo era establecer las bases de un convenio con The Planetary Society, presidida por Carl Sagan, que se redactaría en Pasadena, California (Lemarchand G. A., 2010).

Meses después, The Planetary Society organizó una conferencia sobre búsqueda de vida inteligente en el universo, que se realizó en el Centro de Ciencias de Ontario en Canadá, al que asistieron tanto Lermachand como Colomb. Allí el presidente de The Planetary Society, Carl Sagan, y Colomb, director del IAR, realizaron un acuerdo para la firma de un convenio de cooperación entre el CONICET y la organización que presidía Sagan. Guillermo Lemarchand que estaba presente, tomó una foto de ambos que hoy es histórica (Fig. 9).

Existía interés por parte de la comunidad de SETI de extender su campo de observaciones hacia el hemisferio sur. Tal como relata el mismo Lemarchand, "el convenio implicaba la construcción de una réplica de un analizador espectral de 8,4 millones de canales, con una resolución espectral de 0,05 Hertz por canal, que se encontraba en operación en el radiotelescopio de 26 metros de la Universidad de Harvard". De esta forma con una antena ubicada en el hemisferio norte (el radiotelescopio de la Universidad de Harvard en OAK Ridge, USA) y con otra antena en el hemisferio sur (la perteneciente a uno de los radiotelescopios del IAR), era posible hacer un mapeo en búsqueda de señales de origen extraterrestre, observando constantemente todo el cielo accesible desde la Tierra (3).

Para tal fin, se construyó un analizador espectral que fue denominado META II (acrónimo derivado de "Megachannel Extra-Terrestrial Assay"). Fue necesario que dos ingenieros argentinos del IAR, Eduardo Horrell y Juan Carlos Olade, viajaran a la Universidad de Harvard, donde recibieron entrenamiento de parte del Prof. Paul Horowitz, para la construcción de dicho equipamiento.

El proyecto, denominado como el analizador espectral, fue llamado META II, "hermano" del proyecto META I dirigido por el Prof. Paul Horowitz en EEUU. El proyecto META II se inauguró oficialmente en Argentina en 1990. El analizador espectral fue instalado en el IAR, por lo que la antena II del instituto hizo observaciones del cielo para este proyecto durante casi 20 años (Lemarchand, G. A., 1994; Colomb et al., 1992). Así Argentina se convertía en el único país en desarrollo con un proyecto SETI (Fig. 10).

A partir de la puesta en funcionamiento del proyecto META II, Lemarchand también se dedica a la divulgación del tema a través de distintas publicaciones, tales como el libro "El llamado de las estrellas" (Lemarchand G. A., 1992). El libro contiene 10 capítulos a lo largo de los cuales se describe la historia de los conceptos de la vida más allá de la Tierra, el origen y evolución del universo, de la vida y de la inteligencia, la descripción de la metodología usada para detectar exoplanetas, por esos años en pleno auge, y entre otros temas una discusión sobre el uso de la radioastronomía para detectar señales sobre civilizaciones extraterrestres y el desarrollo de SETI en Argentina.

Luego después de casi 20 años de funcionamiento, el proyecto META II requirió renovación en cuanto a su equipamiento. Para eso era necesario que varios ingenieros viajaran nuevamente a EEUU para construir nuevos analizadores tal como se había realizado en el pasado, pero diversas cuestiones hicieron que el proyecto no pudiera continuarse, incluyendo la incorporación de Guillermo Lemarchand como Consultor principal de la Organización de las Naciones Unidas para la Educación, las Ciencia y la Cultura (UNESCO), quien era el pilar fundamental del proyecto en el país.

Como coronación de esta larga labor, en el año 2009 Lemarchand organiza la "Segunda Escuela de posgrado Iberoamericana de Astrobiología" en Montevideo, Uruguay, a la que asisten investigadores y estudiantes de varios países, incluyendo la autora. Esto además da lugar a una publicación titulada "Astrobiología: del Big Bang a las civilizaciones" (Lemarchand y Tancredi, 2010).

**Nuevos comienzos**

En los últimos años del proyecto SETI en el país, Lemarchand estuvo trabajando en colaboración con dos investigadores, la Dra. Andrea Buccino, estudiante de física en aquel entonces y actual investigadora del CONICET, y otro físico e investigador del CONICET, el Dr. Pablo Mauas, ambos del Instituto de Astronomía y Física del Espacio (IAFE, UBA – CONICET). Junto a ellos, el Lic. Lemarchand, que no sólo era investigador del IAR, sino también investigador del Centro de Estudios Avanzados de la UBA, elaboró una serie de trabajos sobre los efectos de la radiación UV en la habitabilidad de sistemas estelares, en particular focalizado en estrellas de tipo M. Dichos estudios, utilizando abordajes teóricos, lograron una aproximación para la estimar de una zona de habitabilidad UV en planetas extrasolares hipotéticos utilizando para ello cálculos que ponderaban el espectro de acción del ADN (Buccino et al., 2006, 2007).

En el año 2007, tomo conocimiento de estos trabajos, luego de haber obtenido el título de Licenciada en Ciencias Biológicas de la UBA. Por ese entonces también era aficionada a la astronomía con una membresía en el "Club de Astronomía Ingeniero Félix Aguilar", ubicado en Vicente López, provincia de Buenos Aires. A raíz de la lectura de estas publicaciones sobre habitabilidad UV pienso en la posibilidad de que mi formación en biología pudiese ser un aporte para el desarrollo de estas investigaciones.

En función de establecer una colaboración interdisciplinaria entre físicos y biólogos, tomo contacto con el Dr. Mauas del IAFE. Como resultado de esa reunión dispusimos la realización de una tesis doctoral. Así elaboramos un plan de trabajo para llevar a cabo investigaciones incorporando una novedad en el marco de la investigación contemporánea sobre habitabilidad UV que tuvo que ver con la inclusión de experimentos biológicos. Tiempo después, en otra reunión con el Dr. Eduardo Cortón, especialista en bioelectroquímica del Departamento de Química Biológica de la FCEyN-UBA, discutimos nuevos aspectos por lo que fue incorporado como co-director de dicha tesis.

El trabajo de la tesis fue interdisciplinario ya que combinaba tanto aspectos astrofísicos como microbiológicos, y buscó evaluar el efecto de la radiación UV en la habitabilidad planetaria utilizando microorganismos (en particular utilizando arqueas halófilas). A partir de ello se pretendía obtener estimaciones para delimitar una zona de habitabilidad UV en planetas extrasolares hipotéticos. Se obtuvieron resultados preliminares, evaluando la estrella dM EV-Lac (Abrevaya et al., 2008, 2009). El trabajo incluyó adicionalmente otros estudios vinculados a bioelectroquímica aportados por el Dr. Cortón, que estuvo vinculado a un desarrollo en el diseño de unos sensores para detección de vida *in situ* en otros planetas (Abrevaya et al., 2010). Debe tenerse en cuenta que en Argentina no había antecedentes de trabajos similares en ninguno de los abordajes previamente mencionados.

Tres años y medio después, en marzo del año 2011, esta iniciativa comienza a plasmarse durante la finalización de mis estudios de doctorado, donde presento lo que sería primera tesis de astrobiología en la Argentina. Aquel trabajo se tituló: "Efectos de la radiación UV en microorganismos para la elaboración de modelos de habitabilidad y búsqueda de formas de vida simples en planetas solares y extrasolares".

La investigación de la tesis además tuvo un capítulo particular que involucró a científicos de Brasil. El trabajo se inició entre el Dr. Iván Paulino-Lima, biólogo de la UFRJ y la autora, quienes llevaron adelante experimentos junto a un equipo integrado por los doctores Douglas Galante, astrofísico, y Fabio Rodrigues, químico, ambos de la Universidad de Sao Paulo (USP) y por la Dra. Claudia Lage, bióloga de la UFRJ. Así, el grupo de colaboradores de Brasil que también cuenta con experiencia en la realización de estudios interdisciplinarios en condiciones que simulan ambientes extraterrestres, facilitó la ejecución de un trabajo conjunto en el marco de investigaciones sobre litopanspermia. Los experimentos se realizaron en el acelerador sincrotrón que pertenece al Laboratorio Nacional de Luz Sincrotron (LNLS, CNPEM), en Campinas, Brasil y estuvieron vinculados a la exposición de microorganismos extremófilos sometidos a radiación UV en el vacío simulando experimentalmente las condiciones en la órbita baja terrestre durante un viaje interplanetario. Dicho trabajo dio lugar a una publicación en la revista *Astrobiology* y sería el comienzo de una colaboración estrecha en el campo de la Astrobiología entre Argentina y Brasil (Abrevaya et al., 2011; Lage et al., 2012; Rodrigues et al., 2012) (Fig.11).

Luego se sentaron las bases de otra colaboración entre Argentina y Austria, cuando como parte de un postdoctorado en la UBA me surge la posibilidad de una estadía en la Universidad de Graz junto al Dr. Arnold Hanslmeier y los doctores Martin Leitzinger y Petra Odert, del Instituto de Física de dicha universidad. Esta colaboración permitió además el establecimiento de un proyecto de cooperación internacional que fue apoyado y financiado por los ministerios de ciencia de la Argentina y Austria (MINCyT – BMWF, respectivamente, en el período 2012-2015), dirigido por el Dr. Hanslmeier y la autora, titulado: "La influencia de la radiación UV del sol joven sobre la vida".

El esfuerzo por combinar aspectos interdisciplinarios continuó cuando posteriormente regresé a Brasil para realizar un segundo postdoctorado. El trabajo se formalizó en el "Núcleo de Pesquisa em Astrobiologia" del Instituto de Geofísica, Astronomia y Ciencias Atmosféricas de la USP y fue dirigido por el Prof. Dr. Jorge Horvath, astrofísico argentino radicado en Brasil, director del citado Núcleo. Dicho proyecto titulado "Radiación estelar y su influencia sobre la biósfera de cuerpos planetarios" implicó una expansión de los estudios realizados previamente y una vez más combinaba investigación astrobiológica interdisciplinaria que involucraba aspectos microbiológicos y astrofísicos, lo que permitió la consolidación de dos proyectos y una variada cantidad de colaboraciones en el campo de la astrobiología en Brasil y en el exterior. Uno de los proyectos que comenzó a consolidarse fue el llamado "BioSun" cuya semilla inicial se inició en la colaboración con los investigadores de la Universidad de Graz, Austria y que se expandió junto al aporte de los doctores Jorge Horvath astrofísico de la USP y Gustavo Porto de Mello, astrónomo de la UFRJ. Otro trabajo que también comenzó a delinearse en ese entonces fue lo que, años más tarde, se convirtió en el proyecto "EXO-UV", que cuenta con investigadores de Austria, Reino Unido y Brasil.

Durante este postdoctorado, también estreché lazos colaborativos con la Dra María Eugenia Varela, investigadora del CONICET en San Juan, Argentina, quien es experta en cosmoquímica, por lo que su aporte permitió desde ese momento sumar al trabajo interdisciplinario que venía realizando hasta entonces, nuevos aspectos desde el área de la geología que tienen que ver con estudios de litopanspermia.

Luego de una estadía como investigadora visitante en Harvard Smithsonian Center for Astrophysics de la Universidad de Harvard, regresé a la Argentina, para incorporarme como investigadora del CONICET en el Instituto de Astronomía y Física del Espacio, perteneciente a la UBA y al CONICET.

Varias colaboraciones se pusieron nuevamente en marcha en el país, entre ellas se prosiguió con la investigación sobre biosensores para detección de vida que se había iniciado durante la etapa doctoral. Esto llevó a la publicación de un trabajo que fue tapa de la revista *Astrobiology* en el año 2015, la primera que tiene que ver con la labor de científicos argentinos (Figueredo et al., 2015) (Fig. 12).

Tiempo más tarde, luego de 10 años de que iniciara estos estudios interdisciplinarios en Astrobiología y raíz de la colaboración con distintos científicos del país fundé el

"Núcleo Argentino de Investigación en Astrobiología" ("Astrobio.ar") (Fig.13), inspirada en la iniciativa de los colegas de Brasil.

El Núcleo, que actualmente coordino y dirijo, está integrado por investigadores del CONICET y de la Comisión Nacional de Energía Atómica (CNEA), y tiene como objetivo promover la investigación interdisciplinaria en el campo de la astrobiología en Argentina. La formación de los investigadores que lo componen le dan su carácter inter- o multidisciplinario:

-La Dra. María Eugenia Varela, geóloga experta en cosmoquímica, investigadora del CONICET y directora del Instituto de Ciencias Astronómicas de la Tierra y el Espacio (UNSJ - CONICET).

-El Dr. Gerardo Juan M. Luna, astrónomo experto en astrofísica de altas energías. Investigador del CONICET en el Instituto de Astronomía y Física del Espacio (UBA - CONICET).

-El Dr. Oscar Oppezzo, bioquímico experto en fotobiología. Investigador de CNEA y director del grupo de Microbiología Ambiental Aplicada del Departamento de Radiobiología (CNEA).

-La Dra. Nancy López, bióloga experta microbiología ambiental, investigadora del CONICET y directora del Laboratorio de Biotecnología Ambiental y Ecología Bacteriana del IQUIBICEN (FCEyN, UBA - CONICET).

-La Dra. Ana Forte Giacobone, bióloga experta en corrosión microbiológica. Investigadora de la CNEA miembro del Grupo de Microbiología Ambiental Aplicada del Departamento de Radiobiología (CNEA).

-La Dra. Paula Tribelli, bióloga experta en microbiología molecular. Investigadora del CONICET en el Laboratorio de Biotecnología Ambiental y Ecología Bacteriana en el IQUIBICEN, FCEyN – UBA.

Por otro lado, hay varias líneas de investigación que actualmente son llevadas a cabo por el Núcleo. Una descripción de ellas es la que sigue:

1)El estudio de los entornos de radiación en escenarios planetarios y exoplanetarios y sus efectos sobre la habitabilidad. En particular, el estudio del origen de la vida en el Sistema Solar y la posibilidad de la existencia de vida en cuerpos planetarios fuera del Sistema Solar, orbitando diferentes tipos de estrellas. Esto forma parte de dos proyectos internacionales "El proyecto EXO-UV" (EXO-UV Project) y "El proyecto BioSun" (BioSun Project). Ambos proyectos son colaboraciones internacionales y tienen una fuerte contribución tanto de astrofísica como enfoques experimentales de microbiología (ej.: Abrevaya et al., 2020; Abrevaya et al., 2019; Abrevaya y Thomas, 2017; Abrevaya et al., 2015, Abrevaya X.C., 2012a).

2)Simulaciones experimentales de ambientes extraterrestres en condiciones de laboratorio. Esto incluye la radiación, el vacío o la composición atmosférica y la temperatura para recrear las condiciones de escenarios planetarios (por ejemplo, Marte) e interplanetarios (por ejemplo, la órbita baja de la Tierra). Estas simulaciones son parte de diferentes estudios sobre habitabilidad y litopanspermia (una hipótesis que considera a los meteoritos como portadores potenciales de formas de vida microbiana en la transferencia interplanetaria de vida) (ej: Abrevaya et al., 2015; Abrevaya et al., 2016).

3) Microbiología de ambientes extremos como los hipersalinos, en particular centrada en microorganismos como arqueas halófilas o haloarchaea. Dichos microorganismos extremófilos que viven en ambientes con elevadas concentraciones de sal y se consideran modelos de vida extraterrestre potencial, debido a su capacidad para hacer frente a múltiples factores ambientales (radiación, desecación, entre otros) habitan en múltiples tipos de ambientes hipersalinos terrestres que también se consideran análogos de escenarios extraterrestres similares a los que encontramos en cuerpos planetarios del Sistema Solar. Parte de este trabajo tiene como objetivo además revelar las similitudes entre los ambientes hipersalinos de Marte y la Tierra y la posibilidad de considerar a los halófilos como un posible tipo de vida factible de encontrar en otros planetas. Aquí combinamos enfoques desde la geología a la biología (ej.: Abrevaya X.C., 2012b; Abrevaya et al., 2016).

4) Desarrollo de métodos para detectar vida extraterrestre. Trabajamos en el desarrollo de métodos para detectar vida en otros cuerpos planetarios, principalmente en función de sus características metabólicas (ej: Figueredo et al., 2015; Saavedra et al, 2018).

Muchos de los temas desarrollados no sólo tienen la capacidad de generar conocimiento en el campo de la ciencia "básica", sino que también tienen la posibilidad de abrir campos en posibles aplicaciones derivadas, como aquellas relacionadas a la microbiología ambiental y biotecnología, entre otras.

Durante estos años además se ha realizado formación de recursos humanos en el área con la dirección y codirección de tesis (ej: García et al., 2014), y también se han obtenido subsidios orientados a esta área específica, lo que ha ayudado a expandir los conocimientos en este campo.

Debido a que considero como compromiso comunicar la ciencia y a que poseo experiencia en el área de periodismo científico, he impulsado como iniciativa que uno de los objetivos del Núcleo además sea realizar tareas de difusión y comunicación pública de la ciencia en astrobiología y temáticas vinculadas, ya sea generando material a través de iniciativas propias, o prestando servicio a la comunidad a través de entrevistas y notas en medios masivos de comunicación.

**Otras iniciativas relacionadas**

Existe otro esfuerzo reciente, al momento de la escritura de este capítulo, que podríamos ubicar en el terreno de lo que tiene que ver con la búsqueda de vida inteligente en el universo, y que es el Proyecto OTHER (acrónimo de "Otros mundos Tierra, Humanidad y Espacio Remoto").

El proyecto OTHER está liderado por el ex director del Observatorio del Vaticano, el sacerdote católico y astrónomo José Funes, que actualmente es investigador del CONICET en la Universidad Católica de Córdoba. El proyecto se presenta como un "laboratorio de ideas" que intenta dar un enfoque a la búsqueda de otros mundos habitados a través de "la ciencia, la filosofía y la religión", para entre otros temas, discutir y abordar cuestiones como el posible impacto del potencial descubrimiento de una civilización extraterrestre en la concepción filosófica, social y religiosa de nuestra propia civilización (Funes et al., 2016). El grupo de discusión está conformado por varios científicos (en su mayoría astrónomos) y filósofos, y organiza seminarios periódicos para la discusión de estos temas, dentro de los cuales soy invitada a participar.

Otra iniciativa de interés en Argentina, aunque no tiene que ver directamente con la astrobiología, pero está vinculada a la desmitificación de cuestiones vinculadas a la vida extraterrestre y la presencia de objetos voladores no identificados ("OVNIS"), a nivel oficial, es la que implica a la Comisión de Estudio de Fenómenos Aeroespaciales (CEFAE) de la Fuerza Aérea Argentina (FAA). Si bien el organismo fue concebido en el 2011 y creado inicialmente para brindar una explicación a las demandas de la población y de algunos profesionales de aviación ante los crecientes avistamientos de supuestos "OVNIS", recién en el año 2014 con la incorporación del Comodoro (R) Rubén Lianza, es cuando CEFAE adquiere su enfoque científico-educativo. Lianza había trabajado efectivamente en investigación de fenómenos aéreos inusuales durante sus años de actividad, e hizo una propuesta renovadora acerca del enfoque de la Comisión, para tratar el Estudio de Fenómenos Aéreos Inusuales en Organismos del Estado Nacional y fue quien le dio a dicha comisión su "carácter científico" actual. A partir de ese momento el mismo Ministerio de Defensa consideró que no sólo era necesario contar con organismos que dieran una respuesta a la gente, como un servicio público, sino que además se obtendría el doble beneficio de contar con un organismo que generara conocimiento sobre Identificación Aeroespacial, materia escasamente desarrollada en el mundo y que podría ser de suma utilidad tanto para Organismos gubernamentales como para Organizaciones No Gubernamentales o Instituciones Científicas y de difusión.

Así, CEFAE resolvió el ciento por ciento de los casos de avistamientos de objetos y así desarticulaba hipótesis de los "Ovnílogos" (2015-2018). CEFAE no sólo adquirió una invalorable experiencia en un campo muy poco desarrollado sino que también incorporó y tuvo que elaborar sus propias herramientas específicas que le otorgaron a sus informes el adecuado nivel de peritaje científico que exigen los estándares internacionales. Además incorporó asesores externos, plantel al que me he incorporado. Esto surge a partir de Noviembre de 2016 (por Disposición del Secretario Gral. de la FAA) e implica asesores que hubieran acreditado título de grado en carreras científico/técnicas acordes con el enfoque de la Comisión, que claramente excluía cualquier postura orientada hacia la "búsqueda de lo paranormal" (CEFAE, informe institucional, s.f.).

A mediados de 2018, los superiores de la Fuerza Aérea decidieron readecuar su estatus administrativo mediante una Resolución que, si bien dictaminaba que CEFAE cesaba en sus funciones, al mismo tiempo creaba lo que se llamaría el Centro de Identificación Aeroespacial (CIAE), rganismo de mayor nivel organizacional y que retenía todas las atribuciones y responsabilidades de CEFAE, como así también los medios materiales y el personal, de forma tal que ello no le generara al Estado erogaciones adicionales y que entre otras funciones involucra el adiestramiento previo de los futuros oficiales que irían destinados al Comando Aeroespacial del Estado Mayor Conjunto.

Por otro lado, aunque fuera del país, también cabe destacar a un investigador de origen argentino que está vinculado a la temática, el Dr. Gerónimo Villanueva. Luego de realizar un máster en Argentina con una Ingeniería en Telecomunicaciones estudiando la capa de ozono y el vapor de agua en el hemisferio sur (Universidad de Mendoza), en el 2001 Villanueva emigra a Alemania y luego a EEUU donde se convierte en investigador del NASA Goddard Space Flight Center. Como especialista en ciencias planetarias, Villanueva se especializa en la búsqueda de moléculas orgánicas en Marte y en cuerpos helados.

**Algunas reflexiones finales**

En Argentina, el interés por la búsqueda de vida en otros lugares del universo comienza a perfilarse como una ciencia consolidada a partir del proyecto SETI a fines de los años 80. Esto fue desarrollado en nuestro territorio a partir de la iniciativa de astrónomos y astrofísicos, aunque hay varias iniciativas individuales de relevancia que preceden al interés por dicha temática.

Posteriormente las investigaciones en este campo toman otros rumbos que tienen que ver con otra vertiente de la astrobiología, como parte de las investigaciones que estaban siendo desarrolladas en el marco de la astrofísica estelar y adquiere finalmente su matiz actual al incorporar a la biología y abordajes experimentales de laboratorio a partir del año 2007 con el inicio de una tesis doctoral que concluye en el 2011. Dicha vertiente astrobiológica, que tiene que ver con abordajes interdisciplinarios, se consolida a partir de la creación del Núcleo Argentino de Investigación en Astrobiología (Astrobio.ar) actualmente en funcionamiento desde hace algunos años y que se encuentra en expansión impulsando investigación en el área ya sea como parte de colaboraciones nacionales o internacionales.

Otras iniciativas recientes también acompañan dicha expansión, por lo que es de esperar que la astrobiología crezca como ciencia en Argentina en los próximos años, aunque se requiere aún mucho más apoyo a nivel institucional y gubernamental, el cual resulta imprescindible para el crecimiento del área en el país.

**Agradecimientos**

**Referencias**

Abrevaya X.C., Leitzinger M., Oppezzo O.J., Odert P., Patel M., Luna G.J.M., Forte Giacobone A.F., Hanslmeier A. (2020) The UV surface habitability of Proxima b: first experiments revealing probable life survival to stellar flares. MNRAS 494: L69–L74.

Abrevaya X.C., Leitzinger M., Oppezzo O.J., Odert P., Luna G.J.M., Patel M. , Forte-Giacobone A.F., Hanslmeier A. (2019) Towards astrobiological experimental approaches to study planetary UV surface environments. International Astronomical Union Proceedings Series, IAUS345 Origins: From the Protosun to the First Steps of Life 14(S345), 222-226.

Abrevaya, X.C., Thomas, B.C. (2017) Radiation as a constraint for life in the universe. In: Habitability of the Universe before Earth. Eds: Rampelotto, P.H., Seckbach, J., Sharov, A., Gordon, R. Astrobiology Exploring life on Earth and Beyond. Elsevier B.V., Amsterdam. pp 27-46 (Chapter 2).

Abrevaya, X.C. Caneiro A., Horvath J.E., Galante D., Wilberger, D.O., Vega-Castillo J., Rodrigues F., Varela M.E. (2016) Synthesis of Halite Under Martian Simulated Conditions: A Study with Astrobiological Implications. LPSC Contribution No. 1903, p. 2134.

Abrevaya, X.C.; Galante, D.; Nobrega, F.; Tribelli, P.; Rodrigues, F.; Araujo, G.; Gallo, T.; Ribas, I.; Sanz Forcada, J.; Rodler, F.; Porto de Mello, G.F.; Leitzinger, M.; Odert, P.; Hanslmeier, A.; Horvath, J. E. (2015) The use of synchrotron radiation in Astrobiology: Lithopanspermia studies and the Biosun project 25th RAU - Anual Users Meeting LNLS-CNPEM.

Abrevaya, X.C., Hanslmeier, A., Leitzinger, M., Odert, P., Horvath, J.E., Ribas, I., Galante, D., Porto de Mello, G.F. (2014a) The BIOSUN project: an astrobiological approach to study the origin of life. Revista Mexicana de Astronomía y Astrofísica 44:144-145.

Abrevaya, X.C., Hanslmeier, A., Leitzinger,M., Odert, P., Mauas, P.J.D., Buccino, A. P. (2013) UV Radiation of the young Sun and its Implications for life in the Solar System. Central European Astrophysical Bulletin 37: 649-654.

Abrevaya, X.C. (2012a) Astrobiology in Argentina and the Study of Stellar Radiation on Life. BAAA 56: 113-122.

Abrevaya, X.C. (2012b) ́Features and applications of halophilic archaea ́. In: Extremophiles: Sustainable Resources and Biotechnological Implications (Editor: Dr. Om V. Singh), Ed. Wiley. pp 123-158 (Chapter 5)

Abrevaya, X.C., Cortón, E., Mauas, P.J.D (2011a) Flares and Habitability. Proceedings IAU Symposium 286: Comparative magnetic minima: characterizing quiet times in the Sun and stars, Cambridge University Press. S286: 405 - 409.

Abrevaya, X.C., Paulino-Lima, I.G., Galante, D., Rodrigues, F., Cortón E., Mauas P.J.D., de Alencar Santos Lage, C. (2011b) Comparative survival analysis of Deinococcus Radiodurans and the haloarchaea Natrialba magadii and Haloferax volcanii, exposed to vacuum ultraviolet irradiation. Astrobiology 11: 1034-1040.

Abrevaya, X.C.; Mauas P. J. D.; Cortón E. (2010) Microbial fuel cells applied to the metabolically based detection of extraterrestrial life. Astrobiology 10: 965-971.

Bajaja, E. (s.f.) La Radioastronomía en Argentina. En: http://www.cielosur.com/notas_anteriores/bajaja.php

Buccino A.P., Lemarchand G.A., Mauas P.J.D. (2006) UV radiation constraints around the circumstellar habitable zone. Icarus 183: 491-503.

Buccino, A.P., Lemarchand, G.A., Mauas, P.J.D. (2007) UV habitable zones around M stars. Icarus 192: 582-587.

CEFAE, Institucional (s.f.) https://www.faa.mil.ar/mision/cefae.html

Cocconi, G. y Morrison, P. (1959) Searching for Interstellar Communications, Nature 184: 844- 846.

Colomb, F.R., Martín, M.C. y Lemarchand, G.A. (1992) SETI Observational Program in Argentina, Acta Astronautica, 26: 211-212.

de Asúa, M. (2011) "Una gloria silenciosa, dos siglos de ciencia en Argentina". Editorial Libros del Zorzal. 316 pp.

Figueredo, F., Cortón, E., Abrevaya, X.C. (2015) *In situ* search for extraterrestrial life: A Microbial Fuel Cell-Based Sensor for the Detection of Photosynthetic Metabolism. Astrobiology 15: 717-727.

Funes J., Lares M., de los Rios M., Martiarena M., Ahumada A. (2016) OTHER: A multidisciplinary approach to the search for other inhabited worlds. BAAA 59: 1-3.

García, M., Abrevaya, X.C., Gómez, M.N. (2014) Condiciones fisicas de exoplanetas y microorganismos que habitan ambientes extremos. BAAA 56: 407-410.

Lage, C., Dalmaso G., Texeira, L., Bendia A., Paulino Lima, I., Galante, D., Janot-Pacheco E., Abrevaya X.C., Azua-Bustos, A., Pellizari V., Rosado A. (2012) Probing the limits of extremophilic life in extraterrestrial environment simulated experiments. International Journal of Astrobiology 11: 251-256.

Lemarchand G. A. (2010) Una breve historia social de la Asrobiología en Latinoamérica. En: "Astrobiología: del Big Bang a las Civilizaciones" (Lemarchand, G. y Tancredi G., Eds.) Editorial de la Unesco. pp 24-52.

Lemarchand G. A. y Tancredi G. (Eds.) (2010) "Astrobiología: del Big Bang a las Civilizaciones". Editorial de la Unesco. 347 pp.

Lemarchand, G. A. (1986) Jornadas Interdisciplinarias Sobre Vida Inteligente en el Universo, Astrofísica 1: 3-8.

Lemarchand, G. A. (1992) El llamado de las estrellas, Colección Lugar Científico, Buenos Aires, Lugar Editorial.

Mirabel, I.F. (1984) Búsqueda de vida extraterrestre: desarrollos recientes y nuevas perspectivas, Revista Astronómica, No. 230, julio-septiembre.

Reseña Histórica del IAR (s.f.) http://www.iar.unlp.edu.ar/historia.htm

Saavedra, A., Federico, F., Cortón, E., Abrevaya, X. C. (2018) An electrochemical sensing approach for scouting microbial chemolithotrophic metabolisms. Bioelectrochemistry 123:125-136.

Varsavsky, C. M. (1968). Vida en el Universo, Buenos Aires, Carlos Pérez Editor.